\documentclass{aastex631}

\usepackage[T1]{fontenc}

\newcommand{\sfrac}[2]{\mathchoice%
  {\kern0em\raise.5ex\hbox{\the\scriptfont0 #1}\kern-.15em/
    \kern-.15em\lower.25ex\hbox{\the\scriptfont0 #2}}
  {\kern0em\raise.5ex\hbox{\the\scriptfont0 #1}\kern-.15em/
    \kern-.15em\lower.25ex\hbox{\the\scriptfont0 #2}}
  {\kern0em\raise.5ex\hbox{\the\scriptscriptfont0 #1}\kern-.2em/
    \kern-.15em\lower.25ex\hbox{\the\scriptscriptfont0 #2}} {#1\!/#2}}

\newcommand{\Ub}{{\bf{U}}}

\newcommand{\Omegab}{{\bf{\Omega}}}

\newcommand{\epsdot}{\dot{\epsilon}}

\newcommand{\omegadot}{\dot{\omega}}

\newcommand{\C}{$^{12}$C}

\newcommand{\isot}[2]{$^{#2}\mathrm{#1}$}

\newcommand{\castro}{{\sf Castro}}
\newcommand{\amrex}{{\sf AMReX}}

\begin{document}
\title{Comparing Early Evolution of Flames in X-ray Bursts in Two and Three Dimensions}

\shorttitle{Early XRB Flame Evolution}

\author[0000-0001-8401-030X]{Michael Zingale}
\affiliation{Dept.\ of Physics and Astronomy, Stony Brook University,
             Stony Brook, NY 11794-3800}

\author[0000-0001-6191-4285]{Kiran Eiden}
\affiliation{Department of Astronomy,
University of California, Berkeley,
Berkeley, CA 94720-3411, USA}

\author[0000-0003-0439-4556]{Max Katz}
\affiliation{Dept.\ of Physics and Astronomy, Stony Brook University,
             Stony Brook, NY 11794-3800}

\correspondingauthor{Michael Zingale}
\email{michael.zingale@stonybrook.edu}

\begin{abstract}
We explore the early evolution of flame ignition and spreading on the
surface of a neutron star in three-dimensions, in the context of X-ray
bursts.  We look at the nucleosynthesis and morphology of the burning
front and compare to two-dimensional axisymmetric simulations to gauge
how important a full three-dimensional treatment of the flame is for
the early dynamics.  Finally, we discuss the progress toward full-star
resolved flame simulations.\end{abstract}


\section{Introduction}\label{Sec:Introduction}

X-ray bursts (XRBs) result from thermonuclear burning of an accreted
H/He or He layer on a neutron star \citep{galloway:2017}.
Observations of brightness oscillations in the rise of the lightcurve
suggest that the burning begins localized and then spreads across the
neutron star \citep{bhattacharyya:2007}. 
This spreading is inherently
a multi-dimensional phenomenon, requiring hydrodynamic simulations that
resolve the reactive zone, realistic reaction networks, and
domain sizes that capture the scales over which rotation is important in order to
understand the flame propagation and nucleosynthesis.  Detailed
nuclear physics is especially important for understanding the 
rp-process in mixed H/He bursts, as discussed in \citet{rpprocess}.

Here we show a first attempt at modeling the early evolution of a spreading
hotspot in three dimensions.  This builds on our earlier two-dimensional
work \citep{eiden:2020,harpole:2021} that developed our
simulation framework and explored how the acceleration of the burning
front depends on the initial model structure.  As with those
simulations, we use the freely-available \castro\ simulation code
\citep{castro,castro_joss}, with all of the code needed to run these
simulations in our public GitHub repositories.

Our simulations complement the two- and three-dimensional models of flames of
\citet{cavecchi:2013,art-2015-cavecchi-etal,art-2016-cavecchi-etal,Cavecchi2019}
by focusing on resolving the reaction zone of the flame and modeling
the vertical structure hydrodynamically instead of hydrostatically.
We also focus on modeling flames starting from a small hot spot.  This
however means that we are confined to smaller domains, with similar
computational resources, so one of the goals of this study is to
understand how far we can push resolved XRB flame simulations.  The
work by \citet{goodwin:2021} also explores multidimensional evolution,
looking at the thermal transport and how it affects the location where
the hotspot ignites.  Together, all three approaches help build a
comprehensive multidimensional picture of X-ray bursts.

The goal of this paper is to understand the challenges of resolved
three-dimensional hydrodynamical simulation and explore what is
possible with today's resources.  This will inform our follow-on
simulations.

\section{Simulations and Results}\label{Sec:results}

In \citet{eiden:2020}, we developed the simulation methodology in
\castro\ for modeling laterally spreading flames through the accreted
layer on the surface of a neutron star.  For these models,
\castro\ evolves the compressible Euler equations with a Coriolis
force arising from rotation, constant gravity (the plane-parallel
approximation), reactive sources, and thermal diffusion (integrated
explicitly):
\begin{eqnarray}
\frac{\partial \rho}{\partial t} + \nabla \cdot (\rho \Ub) &=& 0 \\
\frac{\partial (\rho X_k)}{\partial t} + \nabla \cdot (\rho \Ub X_k) &=& \rho \omegadot_k \\
\frac{\partial (\rho \Ub)}{\partial t} + \nabla \cdot (\rho \Ub \Ub) + \nabla p &=& \rho {\bf g} - 2 \rho {\bf \Omega} \times \Ub \\
\frac{\partial (\rho E)}{\partial t} + \nabla \cdot (\rho \Ub E + p \Ub) &=& \rho \Ub \cdot {\bf g} + \nabla \cdot k_\mathrm{th} \nabla T + \rho \dot{\epsilon}
\end{eqnarray}
Here, $\rho$ is the mass density, $X_k$ are the mass fractions of the
nuclear species with associated creation rates $\omegadot_k$, $\Ub$
the velocity, $E$ is the specific total energy, $p$ is the pressure,
${\bf g}$ is the gravitational
acceleration, $\Omegab$ is the rotation vector, $k_\mathrm{th}$ is the thermal conductivity, $T$ is the temperature, and $\epsdot$ is the energy release from the reaction network.

\begin{figure}[t]
\centering
\plotone{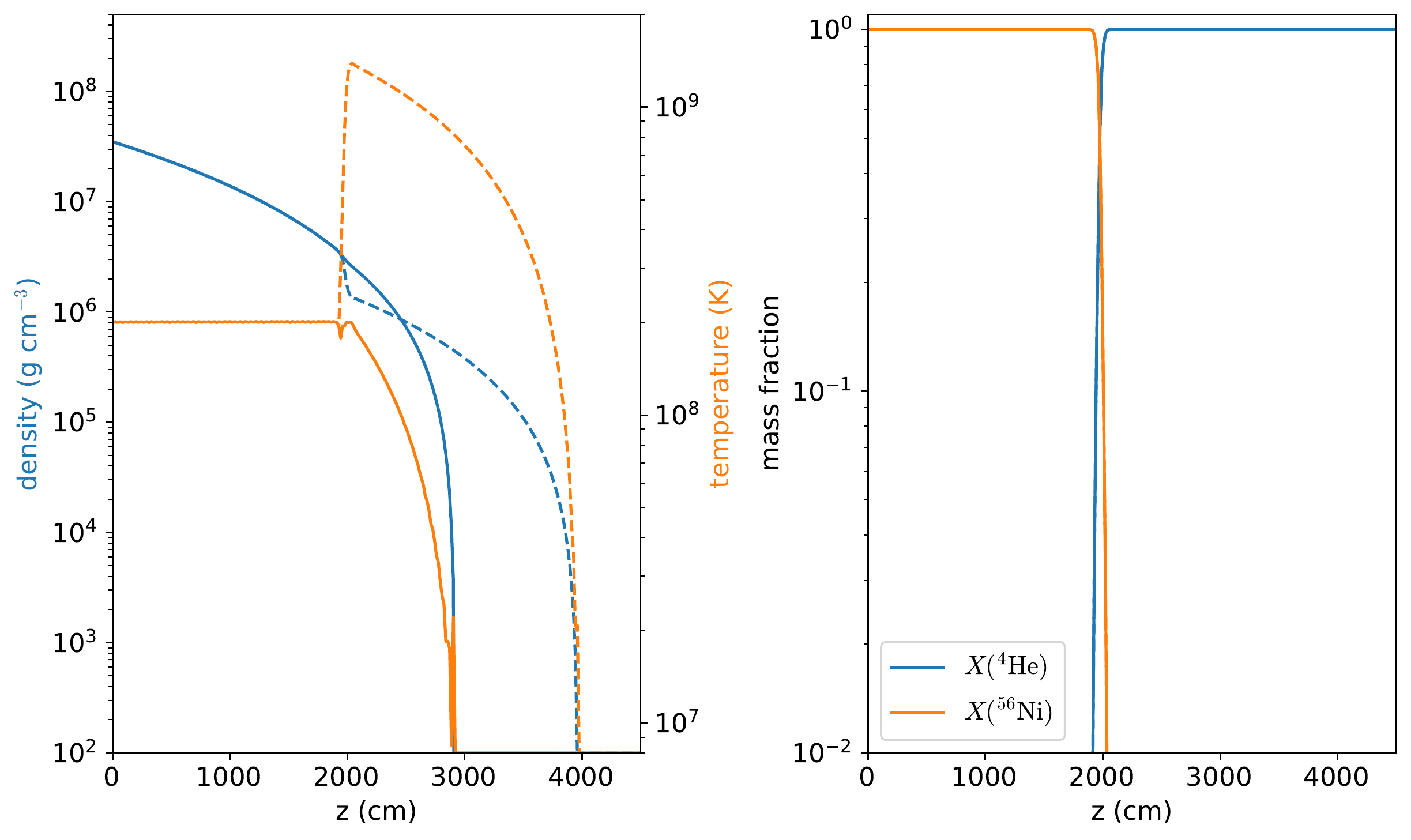}
\caption{\label{fig:initial model} Initial model structure for both
the hot (dashed) and cool (solid) models, showing the density
and temperature structure as well as the initial composition.}
\end{figure}

 We use a general stellar equation of state with
radiation, ions as a perfect gas, and degenerate/relativistic
electrons described in \citet{timmes_swesty:2000} which closes
  the system:
\begin{eqnarray}
E &=& e + \frac{1}{2} |\Ub|^2 \\
p &=& p(\rho, e, X_k) \\
T &=& T(\rho, e, X_k)
\end{eqnarray}
Building on these
first simulations, in \citet{harpole:2021}, we explored several
different rotation rates and initial model thermal structures to
understand how the flame speed depends on the thermal structure of the
envelope.  Both of these papers used a two-dimensional axisymmetric
geometry.  The three-dimensional simulations we present here build on
these works.  We use a neutron star crust temperature of $2\times
10^8~\mathrm{K}$ and rotation rate of 1000~Hz, as followed in
\citet{harpole:2021}.  This model was chosen here since it gives a
clean well-defined flame.  The model is constructed following the same
procedure described in \citet{eiden:2020,harpole:2021}.  We
generate a hot and a cool model, and blend them laterally to setup
an initial perturbation that is in hydrostatic equilibrium.  The
structure of these models is shown
in Figure~\ref{fig:initial model}.

Moving to 3D is very expensive, so to further reduce the computational
expense, we use an anisotropic resolution, with 32 cm resolution
laterally and 16 cm resolution vertically.  This is finer vertically
than done in \citet{harpole:2021}.  Because the hotspot is placed in
the center of the domain, the size of the domain also becomes a
constraint, and as a result, the evolution time we can reach will
become limited by the time it takes the burning to approach the edge
of the domain.  Consequently, we focus here on the early evolution.

\castro\ uses the the \amrex\ adaptive mesh refinement library
\citep{amrex_joss} to manage the discretization and domain
decomposition.  Our 3D simulation used a base grid of $768^2 \times
192$ zones with 2 levels of refinement, the first jumping by a factor
of 4 and the next by a factor of 2.  We subcycle in time, so the
coarser grids take a large timestep than the fine grids.  We used
static mesh refinement, to achieve better load balancing, fully
refining the atmosphere below a height of $3600~\mathrm{cm}$.  The
domain has a size $(1.96608\times 10^5~\mathrm{cm})^2 \times
2.4576\times 10^4~\mathrm{cm}$.  The simulation methodology and
parameters are mostly identical to those used in \citet{harpole:2021}.
We use an unsplit piecewise parabolic method
\citep{ppmunsplit,millercolella:2002} for the advection and operator
(Strang) splitting for the reactions, evolving an internal energy
evolution equation during the burn, as described in
\citet{strang_rnaas}.  These simulations used the 7-isotope He-burning
network described in \citet{iso7}.  This was one of the networks used
in \citet{eiden:2020}.  Thermal diffusion is treated explicitly, using the conductivities from \citet{Timmes00}.
The overall integration strategy is second-order accurate in space and time.  The main difference with the previous
simulations is that we switched from a hydrostatic to reflecting
boundary at the bottom of the domain and used a simple well-balanced
scheme (similar to \citealt{kappeli:2016} but adapted to deal with
characteristic tracing in PPM as described in \citealt{ppm-hse}).
Tests demonstrated that this boundary does a better job than our
previous approach at supporting the atmosphere as the flame propagates
for long times.

The simulations were run on the OLCF Summit supercomputer, on 342 to
1366 nodes, with 6 NVIDIA V100 GPUs per node.  The entire computation
was offloaded to GPUs using the \amrex\ C++ abstraction layer for parallel \texttt{for} loops, as described in
\citet{castro_gpu}.  GPU offloading was critical to being able to
perform these simulations, since we run more than an order of
magnitude faster using all the GPUs on a node as compared to all of
the CPU cores on the node.  Overall, about 250,000 node-hours were used for
the calculation.  The full state of the calculation was output only
every 0.005 s, as each output file is 2.4 TB in size (for single
precision data).  We output a few fields more frequently for
visualization, and global diagnostics were output every timestep.  We
ran a 2D simulation with the same resolution and refinement strategy
to compare with here.

\begin{figure}[t]
\centering
\plotone{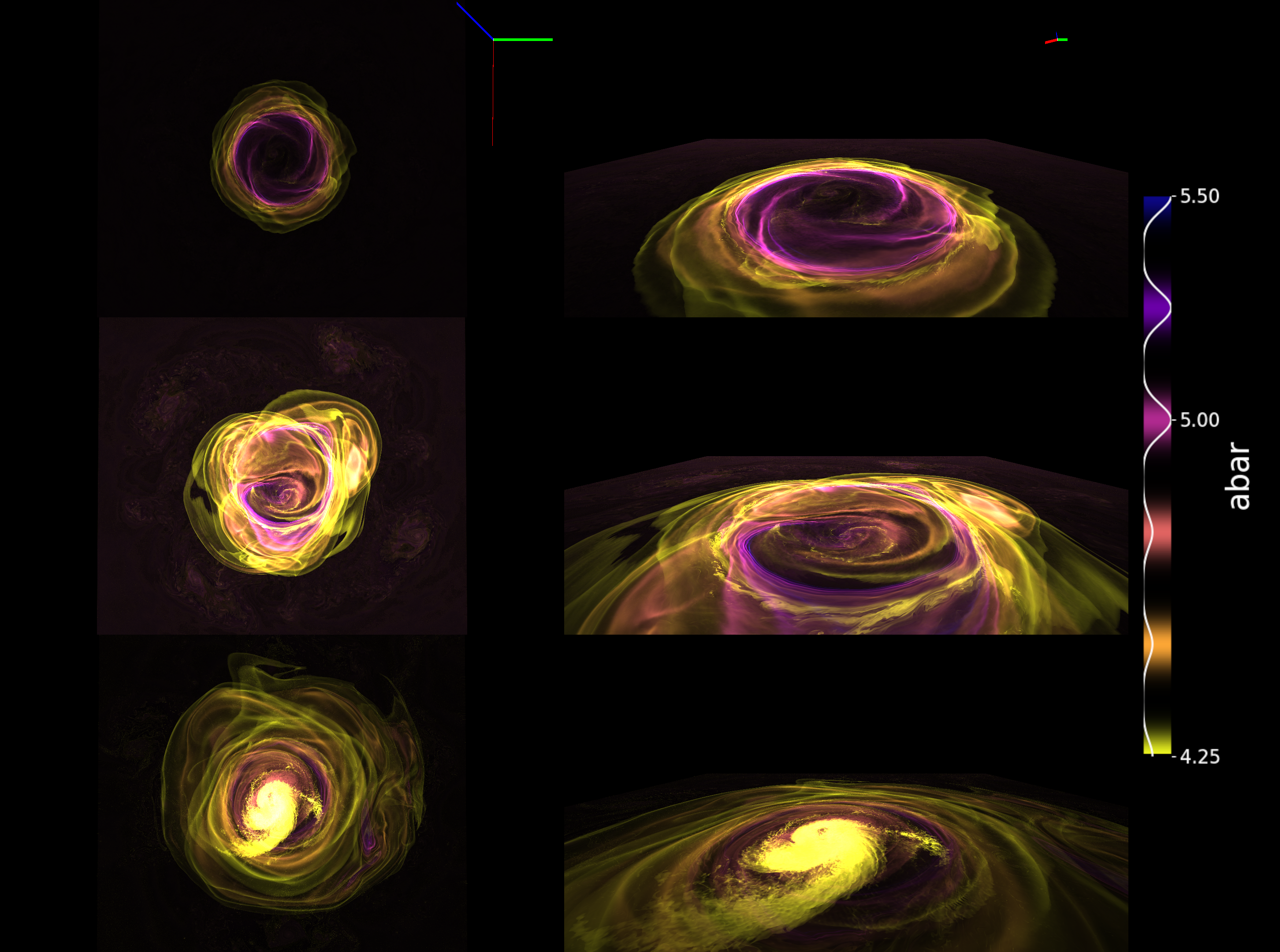}
\caption{\label{fig:vr_abar} Volume rendering of the three-dimensional XRB
simulation showing the mean molecular weight, $\bar{A}$.  Two views
are shown: the left column is viewed from above while the right column
shows it from a shallow angle above the surface.  The panels show the state at 10 ms, 20 ms, and 40 ms (from top to bottom).
The colorbar shows the value of $\bar{A}$ together with the opacity assigned for
that value---this allows us to bring out some structure in the rendering.  A
small triad in the upper right of the top panels shows the orientation,
with red = $x$, green = $y$, and blue = $z$, our vertical direction.}
\end{figure}

Figure~\ref{fig:vr_abar} shows the a three-dimensional simulation via
top-down volume rendering of the mean-molecular weight,
\begin{equation}
\frac{1}{\bar{A}} = \sum_k \frac{X_k}{A_k}
\end{equation}
where $X_k$ is the mass fraction of species $k$ and $A_k$ is the
atomic weight (in mass units).  The structure is shown at 10, 20, and
40~ms of simulation time.  While initially quite symmetric, the axial
symmetry is broken, likely due to roundoff error,  and the flame structure takes on a more tenuous
form.  The panels show the flame spreading considerably through the
domain as time evolves, and the simulation is halted once the flame
nears the boundaries.
\begin{figure*}[t]
\centering
\plotone{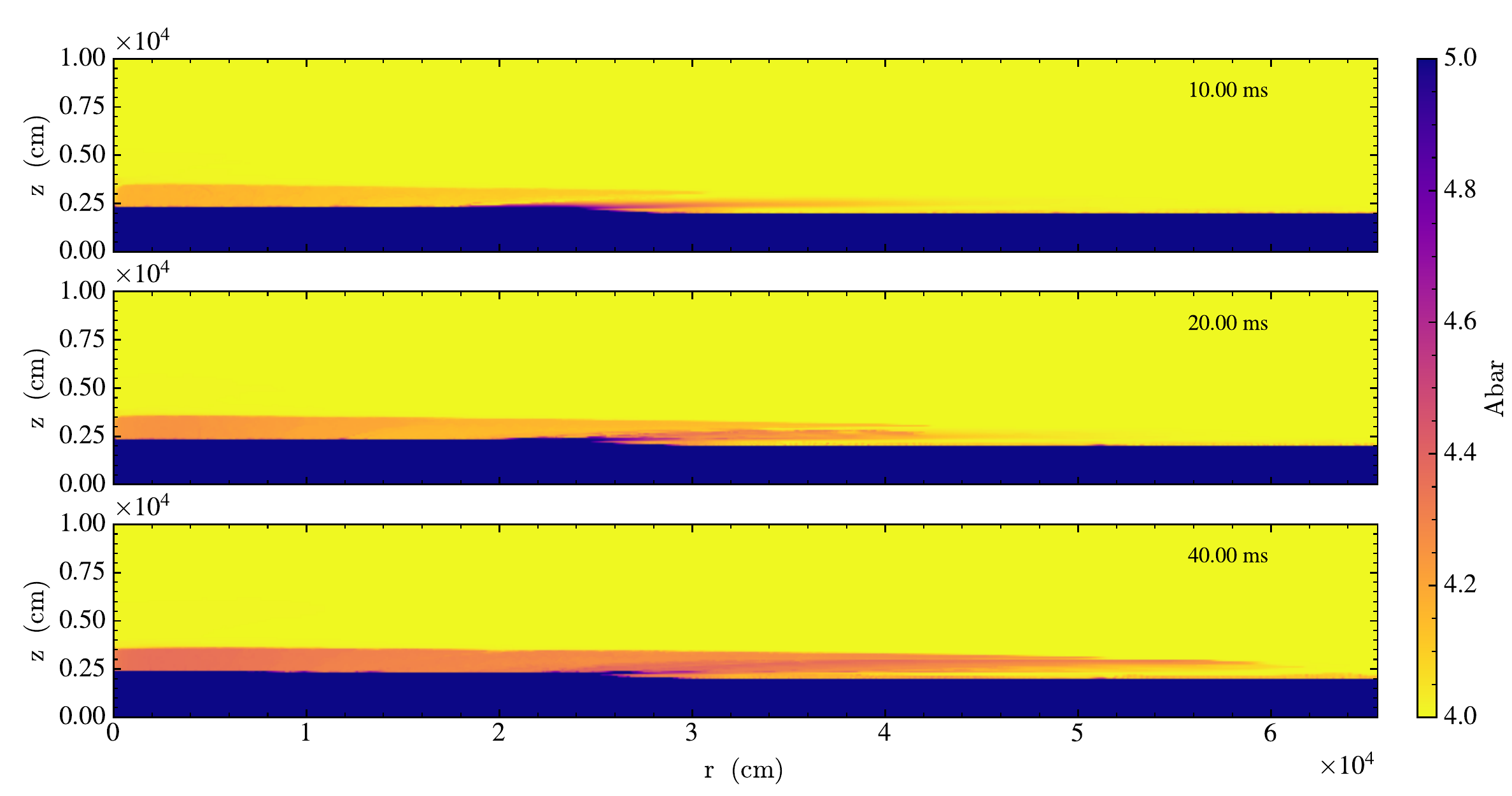}
\caption{\label{fig:2d_abar} Time-sequence of the two-dimensional axisymmetric
simulation using the same setup as the three-dimensional simulation shown
in Figure~\ref{fig:vr_abar}.  The domain is zoomed in, showing only the left
two-thirds of the surface.}
\end{figure*}

\begin{figure*}[t]
\centering
\plotone{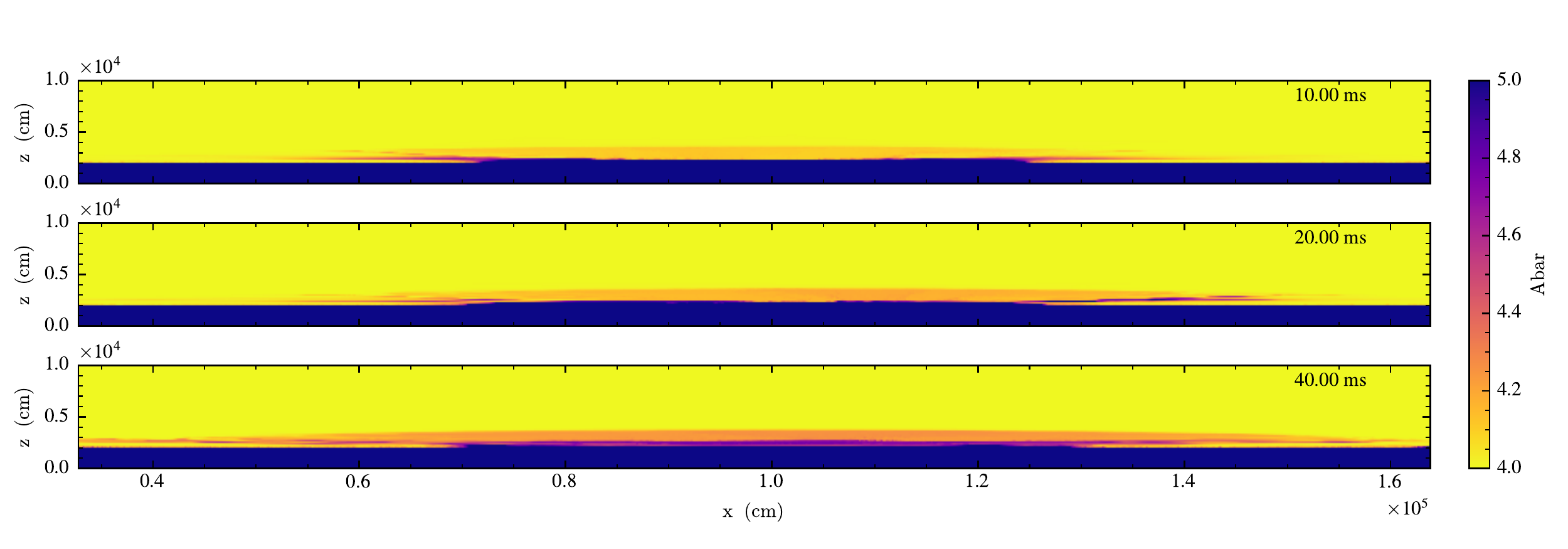}
\caption{\label{fig:3dslice} Vertical slices through the three-dimensional
simulation at the same times as in Figure~\ref{fig:2d_abar}}, showing a region centered on the flame viewed from the side.  The
flame structure is clearly seen. 
\end{figure*}

Figure~\ref{fig:2d_abar} shows the corresponding two-dimensional
axisymmetric simulation.  In the axisymmetric geometry, each zone is
essentially a torus, rotated about the vertical axis.  The initial
model and simulation parameters are identical to the three-dimensional
case.  We see that the overall structure and evolution looks identical
to that reported in \cite{harpole:2021}.  We also see that there is a
well-defined flame in the simulation plane that spreads outward with
time.  A vertical slice through the three-dimensional simulation (in
the $x$-$z$ plane) is shown in Figure~\ref{fig:3dslice} at the final
time, 40~ms.  Aside from the lack of reflection symmetry from the
absence of axisymmetric geometry, it looks very similar to the two-dimensional case.

\begin{figure}[t]
\centering
\plotone{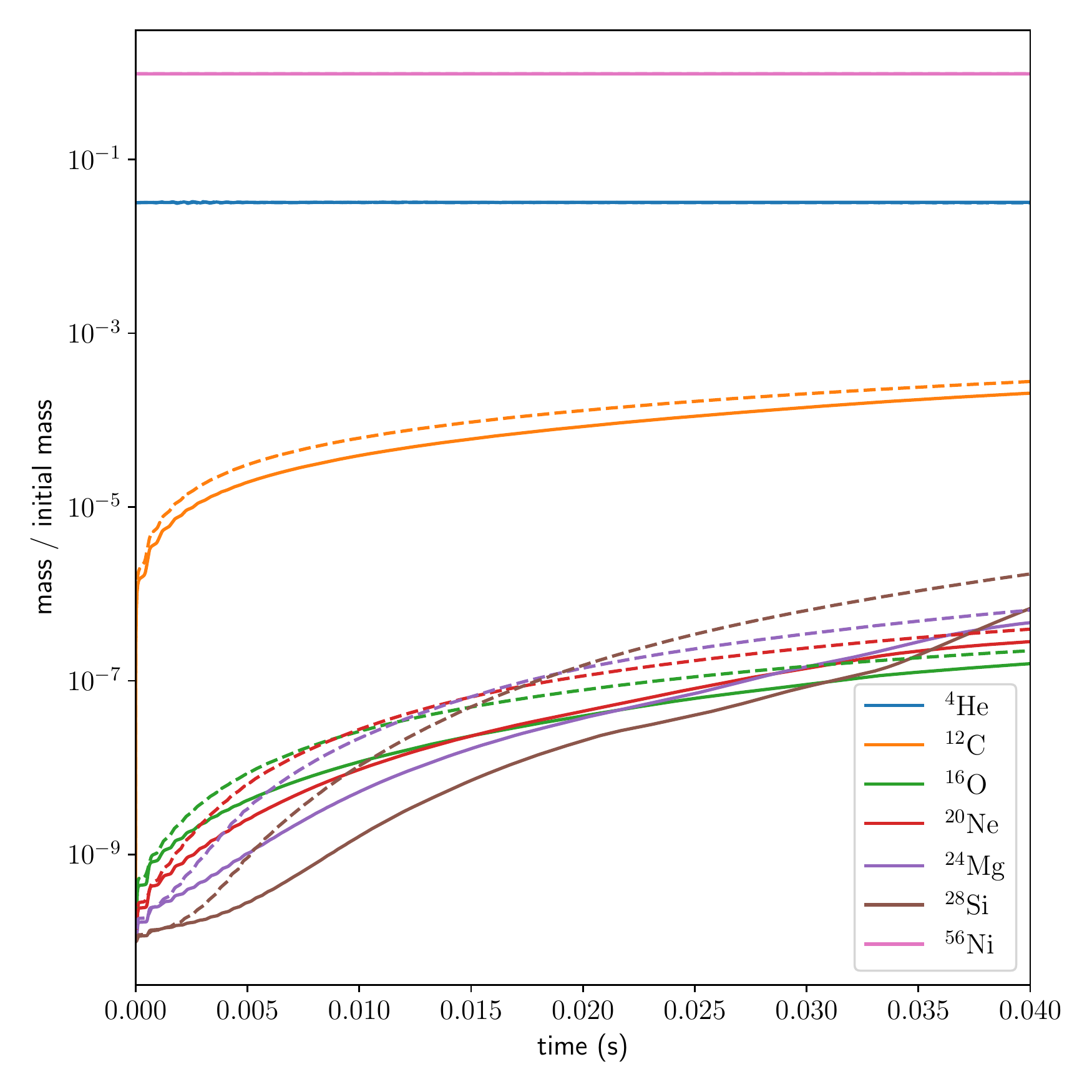}
\caption{\label{fig:mass_plot} Mass of species scaled to total
  mass as a function of time.  The 3D simulation is shown as the solid
  lines and the 2D simulation is shown as the dashed lines.}
\end{figure}

Because the 3D flame does not stay perfectly circular, it is difficult
to define the flame speed using the same method employed in
\citet{eiden:2020}.  Instead, we will look at the mass of the ash
material to assess how quickly the burning is taking place.
Figure~\ref{fig:mass_plot} shows the mass of the different species as
a function of time, for the 2D and 3D calculation.  The axisymmetric
2D calculation has slightly less volume than the 3D domain, since
rotating about the symmetry axis produces a cylindrical domain
inscribed in the 3D domain.  To compensate for this, we scale the
masses by the total mass on the grid.  This plot shows that the
masses of the heavy species, especially \isot{C}{12}, grow quickly
with time.  The differences between the 2D and 3D simulations appear
small---the burning is slightly faster in 2D, but the trends track
very well.  This slight difference is likely because the assumption of
axisymmetry there does not allow for the complex structure we see in
the evolution of composition in the 3D flame.  This strong agreement
suggests that using 2D axisymmetric calculations is a good model for
the early flame evolution.  And since they are so much less
computationally expensive, this will allow us to explore much more
complex reaction networks and understand the nucleosynthesis in more
detail.

\section{Summary}

We have extended our simulation framework to three-dimensions, and
explored the differences between a two-dimensional axisymmetric
simulation and a fully three-dimensional hydrodynamical simulation.
Overall, the early evolution of the flame spreading from a hotspot
behaves similarly in two- and three-dimensions.  As this
three-dimensional simulation pushed the limits of available
computational resources, this result suggests that we can continue to
use two-dimensional axisymmetric simulations to explore the initial
flame structure and spend our computational resources on larger networks
and bigger domains.

These results also suggest that we can do initial explorations of full
star bursts in two-dimensions, perhaps using axisymmetric spherical
coordinates (modeling an annular region in neutron star radius and colatitude).  In this
geometry, point ignition requires starting at the poles, but this
would allow us to see the effect of varying gravity with latitude (due
to centrifugal forces), as well as begin to consider lightcurve modeling.

There is still a role for three-dimensional simulations---the
convective burning in the accreted layer should create turbulence that
the flame will encounter as it propagates.  Although we have shown
that for the flame in a Type Ia supernova, the turbulence from the
era of convection is not strong enough to disrupt the
flame~\citep{wdturb}, the flame in an XRB is considerably slower and
thicker, so it remains to be seen what effect it might have.  This can
be assessed in a different geometry, like a long channel, which would
allow us to trade domain size for resolution to increase the numerical
Reynolds number.  We also will consider higher-order simulation
methodologies which could allow us to capture these effects at lower
resolution.  We've already demonstrated a fully fourth-order accurate
(in space and time) algorithm for coupling hydro, diffusion, and
reactions in \castro~\citep{castro-sdc} that could be used here.  The
main outstanding issue with applying that work to the present problem
is the multilevel time integration.

We are currently exploring mixed H/He bursts as well as the
sensitivity of the flame properties to the size of the reaction
network used.  Both of these initial studies use our two-dimensional
axisymmetric geometry.

\begin{acknowledgements}
\castro\ is open-source and freely available at
\url{http://github.com/AMReX-Astro/Castro}.  The problem setup used
here is available in the git repo as {\tt flame\_wave}. The metadata
describing the build environment and the global diagnostic output
files are available on Zenodo at \citet{xrb_data}.  We thank Alice
Harpole for contributions to the AMReX Astrophysics suite.

The work at Stony Brook was supported by DOE/Office of Nuclear
Physics grant DE-FG02-87ER40317.  This research used resources of the
National Energy Research Scientific Computing Center, a DOE Office of
Science User Facility supported by the Office of Science of the
U.~S.\ Department of Energy under Contract No.\ DE-AC02-05CH11231.
This research was supported by the Exascale Computing Project
(17-SC-20-SC), a collaborative effort of the U.S. Department of Energy
Office of Science and the National Nuclear Security Administration.
This research used resources of the Oak Ridge Leadership Computing
Facility at the Oak Ridge National Laboratory, which is supported by
the Office of Science of the U.S. Department of Energy under Contract
No. DE-AC05-00OR22725, awarded through the DOE INCITE program.  We
thank NVIDIA Corporation for the donation of a Titan X and Titan V GPU
through their academic grant program.  This research has made use of
NASA's Astrophysics Data System Bibliographic Services.
\end{acknowledgements}

\facilities{NERSC, OLCF}

\software{\amrex~\citep{amrex_joss},
          \castro~\citep{castro,castro_joss},
          GCC (\url{https://gcc.gnu.org/}),
          helmeos \citep{timmes_swesty:2000},
          linux (\url{https://www.kernel.org/}),
          matplotlib (\citealt{Hunter:2007}, \url{http://matplotlib.org/}),
          NumPy \citep{numpy,numpy2},
          python (\url{https://www.python.org/}),
          valgrind \citep{valgrind},
          VODE \citep{vode},
          yt \citep{yt}}


\bibliographystyle{aasjournal}
\bibliography{ws}

\end{document}